\documentclass[fleqn,usenatbib]{mnras}

\usepackage[T1]{fontenc}

\DeclareRobustCommand{\VAN}[3]{#2}
\let\VANthebibliography\thebibliography
\def\thebibliography{\DeclareRobustCommand{\VAN}[3]{##3}\VANthebibliography}

\usepackage{graphicx}	
\usepackage{amsmath}	
\usepackage{amssymb}	

\usepackage{newtxtext,newtxmath}

\title[TESS phase curve of WASP-12b]{Phase curve and variability analysis of WASP-12b using TESS photometry}

\author[N. Owens et al.]{
Niall Owens,$^{1}$\thanks{E-mail: nowens09@qub.ac.uk}
E.J.W. de Mooij,$^{1}$
C.A. Watson$^{1}$
M.J. Hooton$^{2}$
\\
$^{1}$Astrophysics Research Centre, Queen’s University Belfast, Belfast BT7 1NN, UK\\
$^{2}$Physikalisches Institut, University of Bern, Gesellschaftsstr. 6, 3012 Bern, Switzerland\\
}

\date{Accepted XXX. Received YYY; in original form ZZZ}

\pubyear{2021}

\begin{document}
\label{firstpage}
\pagerange{\pageref{firstpage}--\pageref{lastpage}}
\maketitle

\begin{abstract}
We analyse Sector 20 TESS photometry of the ultra-hot Jupiter WASP-12b, and extract its phase curve to study the planet's atmospheric properties. We successfully recover the phase curve with an amplitude of 549 $\pm$ 62 ppm, and a secondary eclipse depth of 609$^{+74}_{-73}$ ppm. The peak of the phase curve is shifted by 0.049 $\pm$ 0.015 in phase, implying that the brightest spot in the atmosphere is shifted from the substellar point towards the planet's evening terminator. Assuming zero albedo, the eclipse depth infers a day-side brightness temperature of 3128$^{+64}_{-68}$ K. No significant detection of flux from the night-side is found at 60 $\pm$ 97 ppm, implying a night-side brightness temperature of $<$2529 K (1-$\sigma$). We do not detect any significant variability in the light from the planet over the $\sim$27 days of the TESS observations. Finally, we note that an ephemeris model taking orbital decay into account provides a significantly better fit than a constant-period model. 
\end{abstract}




\section{Introduction}

\begin{figure*}
	\includegraphics[width=\textwidth]{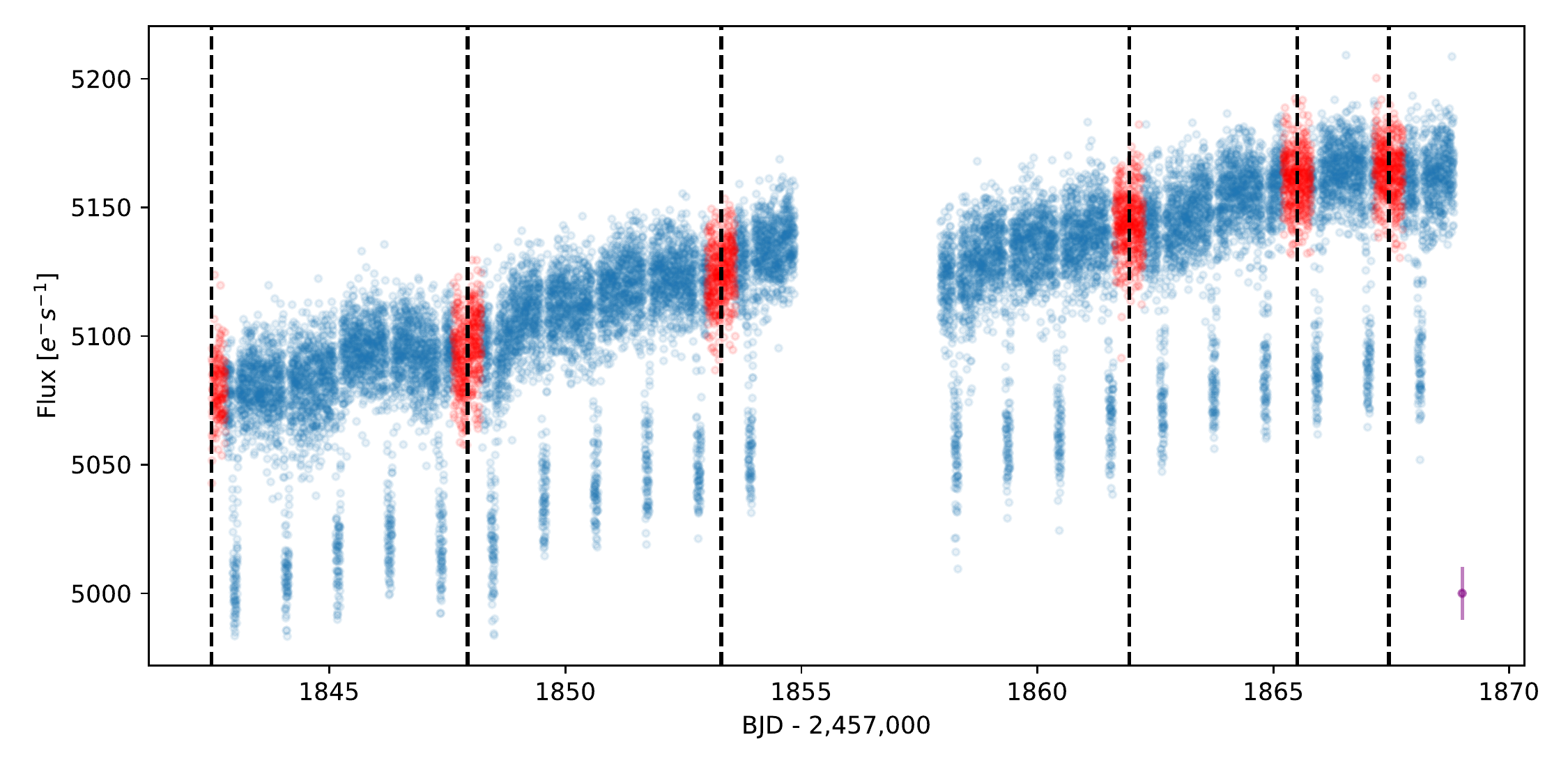}
    \caption{SAP light curve for WASP-12 with flagged points and NaNs removed. Black horizontal lines denote times that were identified as momentum dumps by the data quality flags. Discarded data are shaded in red, spanning $0.6$\,days centred on each momemtum dump. The gap present in the middle is due to a pause to enable data downlink. Clearly present are the 21 transits of WASP-12. A typical errorbar for this dataset is shown in the bottom-right corner.}
    \label{fig:figure1}
\end{figure*}

The atmosphere of a transiting exoplanet can be characterised through observations of its secondary eclipse and phase curve. The depth of the secondary eclipse allows the planetary day-side emission to be measured, providing constraints on the day-side temperature of the atmospheric layers being probed. The day-night contrast can additionally be obtained from the phase curve, as well as effectively mapping the emission of the emitting atmospheric layer as a function of longitude \citep[e.g.][]{knutson2007}. Due to their very nature, hot Jupiters are well suited for this method of characterisation and it has been revealed that their bright spot is often offset in phase from the substellar point \citep[e.g.][]{knutson2007, demooij2009, zhang2018}. This allows the transport of energy within the atmosphere of the planet to be studied.

The spectral energy distribution (SED) for a hot Jupiter peaks in the infrared, making observations in this regime favourable due to the decreased contrast between the planet and its host star \citep[e.g.][]{infra1, bell2}. Due to the strong wavelength-dependence of the planetary flux at optical wavelengths, the contrast between the day- and night-side flux is greater than in the near-infrared. At optical wavelengths, the night-side will appear almost completely dark \citep[e.g][]{snellencorot}. However the planet-star contrast for the day-side is also significantly lower, making detections more challenging. Furthermore, the flux observed in the optical will constitute both thermally emitted- and reflected starlight that - when combined with multi-wavelength studies - can effectively constrain the planet's geometric albedo. One very interesting advantage of this wavelength regime is that it is very sensitive to temperature changes within the layer of the atmosphere being probed, potentially revealing temporal-evolving features. This was the case in \citet{armstrong} who observed the peak emission of the hot Jupiter HAT-P-7b shifting between the morning and evening hemispheres, which could be due to variable cloud coverage.

In this Letter we investigate the phase-curve of the ultra-hot Jupiter WASP-12b \citep{wasp12b}, which orbits a G0-type star every 1.09 days. This makes it one of the most heavily irradiated exoplanets known, with an expected day-side temperature of $\sim$3,000K. Its extreme temperatures have allowed for its secondary eclipse to be observed throughout the infrared \citep{croll, campo, zhao, crossfield, cowan, stevenson, bell2} as well as the optical \citep{lopes, foehring, hooton, vonessen}. Additionally, the geometric albedo for WASP-12b has been measured to be very low ($A_{\mathrm{g}}$ < 0.064) at optical and near-ultraviolet wavelengths \citep{bell} suggesting the planet to be one of the darkest known to date. Of particular interest is the observed variability in reported secondary depths within the $z'$-band; the secondary eclipse observed in \citet{lopes} and \citet{foehring} were conducted just 2 months apart yet disagree in their reported depths by over 3$\sigma$. Further secondary eclipse depths reported by \citet{hooton} in the $i'$-band are also discrepant on the order of 3$\sigma$ and were conducted a year apart. Such a discrepancy corresponds to variations of around 400K which may be the result of atmospheric variability, such as storms. Furthermore, WASP-12b marks the first planet where orbital decay due to tidal dissipation has been confidently observed \citep{orbitdecay}, see also recent work by \cite{turner2020}.

Here we present the first full optical phase curve of WASP-12b as observed by \textit{TESS}. In Section \ref{observations}, we discuss the observations and preparation of the data for analysis; in Section \ref{analysis} we outline our detrending strategy as well as our light curve model; in Section \ref{discussion} we present our results and discuss their implications, including a search for variability within the data as well as orbital decay; and finally we present our conclusions in Section \ref{conclusion}.

\section{\textit{TESS} Observations}\label{observations}
The \textit{TESS} spacecraft observed WASP-12 (TIC 86396382) at 2-minute cadence during its Sector 20 observations from 2019 December 24 to 2020 January 20, over 2 full \textit{TESS} orbits.

The 2-minute data collected by \textit{TESS} is processed by the TESS Science Processing Operations Center \citep[SPOC,][]{spoc} and produces the data products that we use in this analysis. These files consist of both simple aperture photometry (SAP) and presearch data conditioning simple aperture photometry (PDCSAP) light curves. The SAP light curve is produced using an optimum aperture mask as determined by \textit{TESS}. This data has been calibrated, background-subtracted, and has had cosmic ray events removed but has not been detrended or had any other systematic error correction techniques applied. The PDC light curve is treated similarly but has been thoroughly detrended using SPOC algorithms, and makes use of cotrending basis vectors (CBVs) to find common trends within the ensemble flux data that are then subsequently removed from the target flux-series.

Before fitting and performing correction for instrumental effects, all flagged and NaN points were removed. Flagged points also include reaction wheel desaturation events (momentum dumps). During these events, fine pointing is lost, compromising the quality of the data both leading up to- and after the event. This results in large discontinuities within the data about these points and any attempts to detrend such events prove to be challenging. We therefore decided to conservatively remove $0.6$\,days of data centred on each momentum dump due to the severe unpredictable ramp-like fluctuations that take place. Furthermore, the published 1$\sigma$ errors are found to be significantly overestimated due to an issue in the \textit{TESS} analysis pipeline (TESS Science Support Center priv. comm.), so we opt to calculate the spread in the residuals between the data and best-fit model (see Sect.~\ref{sec:phasecurve}) and then scale the errors appropriately.

In total we analyse $14,243$\,points, comprising 21 transits and 20 secondary eclipses. The lightcurve can be seen in Fig.~\ref{fig:figure1}.

\section{Analysis}\label{analysis}
\subsection{\textbf{Light curve model}}\label{sec:phasecurve}
The light curve model was generated using a near-identical model to that used by \citet{esteves2013, esteves2015} where we assume a circular orbit. The model accounts for the transit $F_\mathrm{transit}$, the secondary eclipse $F_\mathrm{ecl}$, the phase-curve modulation $F_\mathrm{p}$, and the effect due to ellipsoidal distortions and Doppler beaming $F_{\mathrm{m}}$. Each term is a function of the planet's phase $\phi$, with transit and secondary eclipse at a phase of $0$ and $0.5$, respectively. The entire light curve is modelled:
\begin{equation}
    F_{\mathrm{transit}}(\phi) \cdot (F_{\mathrm{m}}(\phi)+1) + F_{\mathrm{ecl}}(\phi) + F_{\mathrm{p}}(\phi + \theta)
	\label{eq:phasecurve}
\end{equation}

\noindent where $\theta$ is the brightness offset in phase from the planet's substellar point. We use the transit model $F_\mathrm{transit}$ as formulated by \citet{mandel} and use the Python package \texttt{batman} \citep{batman} to calculate the transit light curve for the WASP-12 system. Note that we use a quadratic limb-darkening law in these solutions, and in our fits we constrain the limb-darkening coefficients (LDCs) by fitting for $q_{1}$ and $q_{2}$ as advocated by \citet{ldc}. From the transit we can determine the planet-to-star radius ratio $R_{\mathrm{p}}/R_{*}$, the scaled planetary semi-major axis $a/R_*$, and the orbital inclination $i$. While the period $P$ and midtransit time $t_{0}$ can be determined as well, we opt to use the ephemeris as described in \citet{orbitdecay} as the short baseline of the TESS data would not allow the orbital decay of WASP-12b to be directly measured.

The secondary eclipse $F_\mathrm{ecl}$ is modelled by normalising the transit solution assuming a uniformly bright source and scaling appropriately. The planet's apparent surface is modelled as a Lambert sphere and is described by:
\begin{equation}
    F_{\mathrm{p}}(\phi) = A_{\mathrm{p}} \frac{\sin{z(\phi)}+(\pi - z(\phi)) \cos{z(\phi)}}{\pi}
	\label{eq:phase1}
\end{equation}
\noindent where $A_{\mathrm{p}}$ is the peak brightness amplitude, and $z(\phi)$ is defined as:
\begin{equation}
    \cos{z(\phi)} = - \sin{i} \cos{(2\pi[\phi+\theta])}
	\label{eq:phase2}
\end{equation}
The final parameter we model and fit is $F_{\mathrm{m}}$, which accounts for Doppler boosting of the system as well as ellipsoidal variations due to tidal effects on the star by the planet. As both of these effects are dependent upon the planetary mass $M_{p}$, they can both be expressed as a single equation:
\begin{eqnarray}
    \begin{aligned}
    F_{m} &= M_{p} \Bigg\{ \bigg(\frac{2 \pi G}{P}\bigg)^{1/3} \frac{\alpha_{d} \sin{i}}{cM_{*}^{2/3}}f_{d} - \frac{\alpha_{2} \sin^{2}{i}}{M_{*}} f_{e} \Bigg\}\\
    f_{e} &= \bigg(\frac{a}{R_{*}}\bigg)^{-3}\cos{(2 \cdot 2\pi \phi)} + \bigg(\frac{a}{R_{*}}\bigg)^{-4}f_{1}\cos{(2\pi \phi)} \\
    &+ \bigg(\frac{a}{R_{*}}\bigg)^{-4}f_{2}\cos{(3 \cdot 2\pi \phi)} \\
    f_{d} &= \sin{2\pi\phi}
    \end{aligned}
\end{eqnarray}
where $G$ is the gravitational constant, $c$ is the speed of light, and $\alpha_{d}$ is the photon-weighted bandpass-integrated beaming factor. $\alpha_{d}$ is calculated as outlined in \citet{doppler1, doppler2}. $\alpha_{2}$ is a function of the limb- and gravity darkening coefficients within the \textit{TESS} waveband for WASP-12 \citep{claret}. $f_{d}$ and $f_{e}$ describe the phase-dependent modulations for Doppler boosting and ellipsoidal variations, respectively, with $f_{1}$ and $f_{2}$ being constants determining the amplitude of the higher-order ellipsoidal terms and are dependent on the limb- and gravity darkening coefficients for the star. The calculation of these terms as well as that of $\alpha_{2}$ are outlined in \citet{esteves2013}. We note that $F_{\mathrm{m}}$ does not constitute part of the planetary phase curve but is due to the star influencing the overall light curve. While these features are detected, the planetary mass still remains unconstrained; we therefore opt to use Gaussian priors (see Table \ref{tab:results}).

\begin{table}
	\centering
	\caption{Best-fit parameters, and derived parameters, from our MCMC fit for WASP-12b.}
	\label{tab:results}
	\begin{tabular}{lccr}
		\hline\hline
		Parameter & Prior & Best-fit\\
		\hline
		\multicolumn{3}{c}{Orbital parameters} \\
		\hline
		$P$ (days) & fixed & 1.091420107(42) $^a$\\
		\rule{0pt}{3.5ex}$t_{\mathrm{0}}$ (BJD) & fixed & 2456305.455809(32) $^a$\\
		\rule{0pt}{3.5ex}$dP/dN$ & fixed & -10.04(69)$\times$10$^{-10}$ $^a$\\
		\rule{0pt}{3.5ex}$a/R_{*}$ & $\mathcal{N}$(3.03, 0.034) $^b$ & 3.038 $\pm$ 0.022\\
		\rule{0pt}{3.5ex}$i$ (degrees) & $\mathcal{N}$(83.37, 0.7) $^b$ & 83.21$^{+0.51}_{-0.48}$\\
		\hline
		\multicolumn{3}{c}{Transit parameters} \\
		\hline
		$R_{\mathrm{p}}/R_{*}$ & - & 0.1174 $\pm$ 0.0007\\
	    \rule{0pt}{3.5ex}$u_{1}$ & - & 0.225$^{+0.177}_{-0.119}$\\
	    \rule{0pt}{3.5ex}$u_{2}$ & - & 0.282$^{+0.194}_{-0.136}$\\
	    \hline
	    \multicolumn{3}{c}{Phase curve} \\
	    \hline
	    $F_{\mathrm{ecl}}$ (ppm) & - & 609$^{+74}_{-73}$\\
	    \rule{0pt}{3.5ex}$A_{\mathrm{p}}$ (ppm) & - & 549 $\pm$ 62\\
	    \rule{0pt}{3.5ex}$\theta$ (phase) & - & 0.049 $\pm$ 0.015\\
	    \rule{0pt}{3.5ex}$M_{\mathrm{p}}$ (M$_{\mathrm{J}}$) & $\mathcal{N}$(1.47, 0.076) $^b$ & 1.48 $\pm$ 0.08\\
	    \rule{0pt}{3.5ex}$F_{\mathrm{n}}$ (ppm) & - & 60 $\pm$ 97\\
		\hline
		\multicolumn{3}{c}{Derived} \\
		\hline
		$A_{\mathrm{g, ecl}}$ & - & 0.408 $\pm$ 0.050\\
		\rule{0pt}{3.5ex}$T_{\mathrm{B,day}}$ (K) & - & 3128$^{+64}_{-68}$\\
		\rule{0pt}{3.5ex}$T_{\mathrm{B,night}}$ (K) & - & <2529 (1-$\sigma$)\\
		\rule{0pt}{3.5ex}$A_{\mathrm{g, max}}$ & - & no solution\\
		\rule{0pt}{3.5ex}$T_{\mathrm{max}}$ (K) & - & no solution\\
		\rule{0pt}{3.5ex}$A_{\mathrm{g, hom}}$ & - & 0.385$^{+0.060}_{-0.061}$\\
		\rule{0pt}{3.5ex}$T_{\mathrm{hom}}$ (K) & - & 2064$^{+105}_{-120}$\\
		\rule{0pt}{3.5ex}$f_{\mathrm{lim}}$ & - & 0.556 $\pm$ 0.047\\
		\hline
		\multicolumn{3}{l}{\footnotesize$^a$ Orbital decay model from \citet{orbitdecay}.} \\
		\multicolumn{3}{l}{\footnotesize$^b$ Gaussian priors from \citet{collinsmass}.}
	\end{tabular}
	
\end{table}

\subsection{Detrending strategy and joint-fit}
In detrending the data, we make use of CBVs to fit the long- and short term systematic trends within the data to minimise the impact other detrending techniques may have on the phase curve. The CBVs come ordered in terms of significance, and we progressively add more CBVs (of lower significance) and fit the data, calculating the BIC for each fit. We select the model that has the BIC minimised, and this occurs when we include 9 CBVs. We opt to fit this number of CBVs in addition to our phase curve model. We note that beyond 2 CBVs, there is no significant difference in the derived planet's parameters.

The CBV function we utilise is of the form:
\begin{equation}
    CBVs =  a_{0} + \sum_{k=1}^{8} a_{k} C_{k}(t)
\end{equation}
where $C$ represents a given CBV as a function of time $t$, and $a$ represents the coefficient or weighting given to that CBV. The full model fit is therefore given as:
\begin{equation}
     [F_{\mathrm{transit}}(\phi) \cdot (F_{\mathrm{m}}(\phi)+1) + F_{\mathrm{ecl}}(\phi) + F_{\mathrm{p}}(\phi + \theta)] \times CBVs
\end{equation}
We perform our fit using \texttt{emcee} \citep{emcee} and use 64 walkers and 15,625 steps for a total of 1,000,000 links with 29 free parameters. Table \ref{tab:results} outlines the Gaussian priors used.

For completeness, we also take the traditional approach of fitting a polynomial to each orbit to remove slow trends, and additionally present the PDCSAP flux. Both analyses can be found in Appendix \ref{app:1}.

\begin{figure*}
	\includegraphics[width=\textwidth]{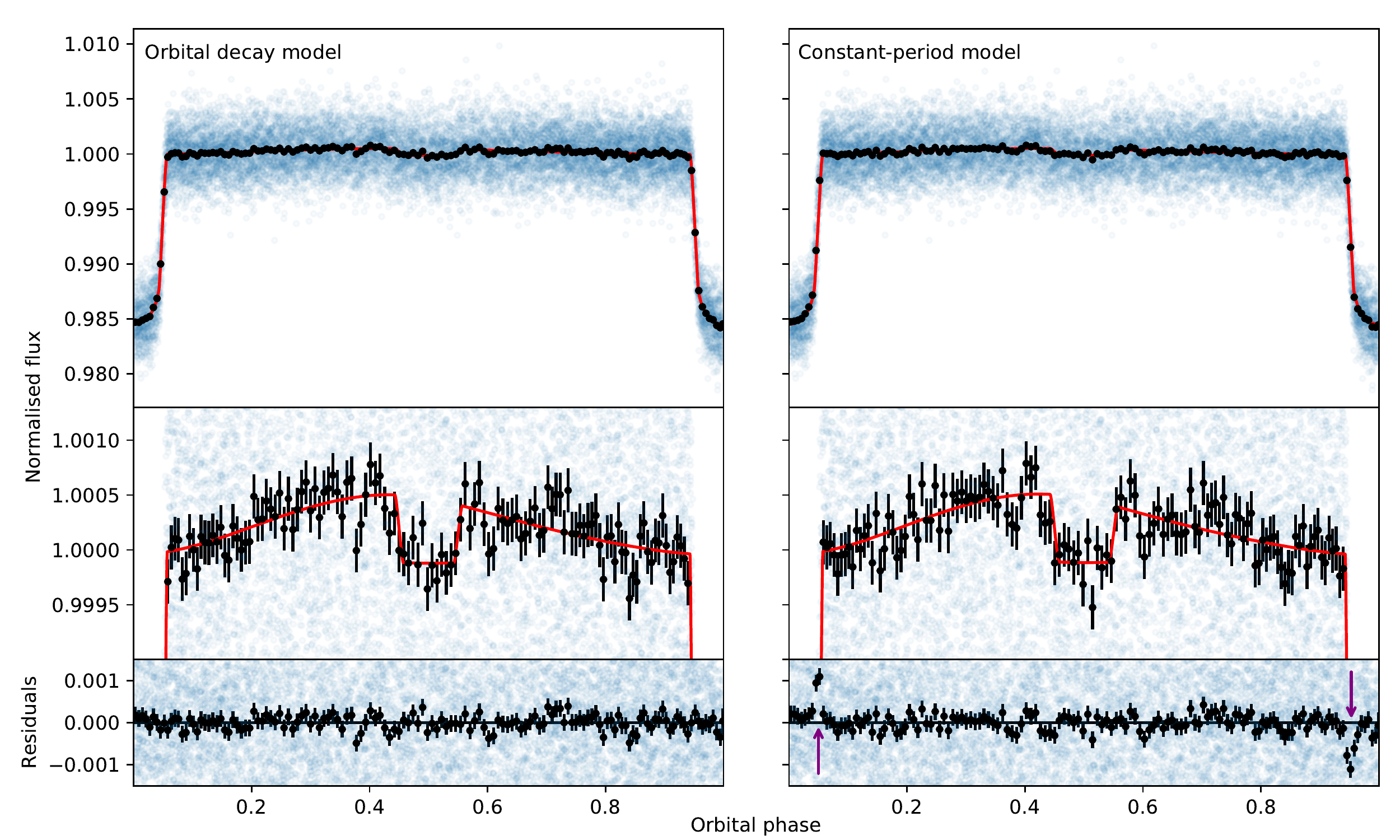}
    \caption{The phase curve of WASP-12b as observed by \textit{TESS}. Unbinned data plotted in blue, and points in black represent 100-point bins and associated uncertanties. \textit{Left}: The decaying ephermeris model used here is outlined by \citet{orbitdecay}. \textit{Right}: Using a non-decaying, constant-period orbit produces very notable artefacts (purple arrows) in the residuals. The phase curve is best fit with a decaying ephemeris, strongly indicating that WASP-12b's orbit is decaying.}
    \label{fig:figure2}
\end{figure*}

\section{Discussion}\label{discussion}

The results of our MCMC  fit are presented in Table \ref{tab:results} and Figure \ref{fig:figure2}. In addition, we performed a residual permutation bootstrap analysis to investigate the impact of correlated noise on the uncertainties. We found that this resulted in smaller uncertainties on the best fit parameters, indicating that the impact of systematic effects was minor. We therefore adopt the uncertainties from the MCMC fit. The results of the fits to each of the individual \textit{TESS} orbits are provided and compared in Appendix \ref{app:1}, we see no significant differences between the key parameters.

\subsection{Brightness temperatures, and geometric albedos}
We can calculate the geometric albedo for WASP-12b if we assume no thermal contribution:
\begin{equation}
    F_{\mathrm{ecl}} = A_{\mathrm{g, ecl}}\bigg(\frac{R_{\mathrm{p}}}{a}\bigg)^{2}.
\end{equation}
We find for WASP-12 that $A_{\mathrm{g,ecl}}<0.458$ (1$\sigma$).
If we assume a bond albedo $A_{\mathrm{B}}$ of 0, then we can determine the planet's equilibrium temperature - which assumes no reflective component - to be \citep{seager}:
\begin{equation}
    T_{\mathrm{eq}} = T_{*}\bigg(\frac{a}{R_{*}}\bigg)^{-1/2}f^{1/4}
\end{equation}
where $f$ is a factor that defines the two limiting cases: homogeneous re-distribution ($f=1/4$) designated $T_{\mathrm{eq,hom}}$, and instant re-radiation ($f=2/3)$ as $T_{\mathrm{eq,max}}$, and $T_{*}$ and $R_{*}$ are the stellar temperature and radius, respectively. These cases are indicative of the efficiency of the atmosphere to absorb and distribute the incoming radiation, and is dependent on the composition of the planetary atmosphere. For WASP-12b, we find $T_{\mathrm{eq,max}}$ = 3275~K, and $T_{\mathrm{eq,hom}}$ = 2560~K. In addition, we calculate the brightness temperatures for both the day- and night-side of WASP-12b, following \cite{esteves2013}. We take the TESS bandpass into account, and use stellar models from the BT-Settl grid \citep{allard} for stellar parameters matching that of WASP-12 as determined by \citet{stellarparams}. These brightness temperatures can be found in Tab.~\ref{tab:results}. We note that, given the bandpass of \textit{TESS}, the eclipse depth will be a combination of both reflected light and thermal emission. As in \cite{esteves2013}, we try to self-consistently solve for the geometric albedo by assuming isotropic scattering (A$_{\mathrm{B}}$ = $\frac{3}{2}A_{\mathrm{g}}$) for two limiting cases in terms of heat distribution throughout the atmosphere, homogeneous redistribution and instant re-radiation.

For a homogeneous temperature the day-side eclipse depth would indicate an albedo of $A_{\mathrm{g, hom}}$ = 0.385$^{+0.060}_{-0.061}$ and an atmospheric temperature, $T_{\mathrm{hom}}$ =  2064$^{+105}_{-120}$K. This would require a much higher albedo than is expected for hot-Jupiters and we consider such a result unrealistic. Furthermore, this is inconsistent with the eclipse depths measured at longer wavelengths \citep[e.g.][]{croll}. 

In the instant re-radiation limit, there is no simple self-consistent model that can explain the emission in the TESS band, suggesting that there must be some energy transport towards the night-side of WASP-12b. Fixing the geometric albedo to zero, we find an upper limit on $f$ at 0.556 $\pm$ 0.047 with a temperature $T$=3128~K. We also attempted to perform a joint fit for our \textit{TESS} data together with \textit{Spitzer} data from \cite{bell2}, and find $A_{\mathrm{g}}$ = 0.32 $\pm$ 0.22 and $T_{\mathrm{day}}$ = 2510 $\pm$ 134~K, which is consistent with the trend between these parameters found by \cite{wongetal}.

The night-side flux $F_{\mathrm{n}}$ is consistent with zero, indicating a large day-night contrast as expected from these ultra-hot Jupiters \citep[e.g.][]{contrast1, contrast3}. This is supported by the low-energy redistribution efficiency found above.

\subsection{Peak brightness offset and variability}
We find that the peak planetary brightness is shifted towards the evening terminator by 0.049 $\pm$ 0.015 in phase , which is typically expected from such planets due to equatorial super-rotation \citep{superrotation}. \citet{bell2} measured both a morning and evening offset in the infrared at 3.6$\mu m$, measuring 0.091 $\pm$ 0.017 degrees in 2010 and -0.038 $\pm$ 0.011 degrees in 2013\footnote{\citet{bell2} uses a different convention and
measure their offsets \textit{after} eclipse. We have converted their values to our phase and sign convention.}. Cloud decks could also cause such observed variability but given WASP-12b's derived temperatures, clouds are unlikely to form. Additionally, WASP-12b has been found to be undergoing orbital decay \citep{orbitdecay} and possibly mass-loss \citep{bell2}, which may also play a role in observed variability. However, we observe no significant difference between either \textit{TESS} orbit (see Appendix~\ref{app:1}).

The secondary eclipses measured for each orbit are consistent within their uncertainties, and so we can not claim any variability. This may be due to the timescales involved. While we can't directly compare our eclipse depths with previous studies as they use different bandpasses, our results lie between the two observed eclipse depths in \citet{hooton}, and are consistent at 2$\sigma$.

\section{Conclusions}\label{conclusion}
We present the full optical phase curve of WASP-12b as observed by \textit{TESS} during year 2 of its mission. With effective detrending and modelling, we are able to derive atmospheric temperatures as well as geometric albedos for limiting cases with regards to heat/energy distribution throughout the planet's atmosphere. We find our results to be consistent with previous measurements, with WASP-12b having a very low albedo and its peak brightness offset towards the evening terminator. We also find that the series is better fit when using an ephemeris model that incorporates the orbital decay outlined in \citet{orbitdecay}. Lastly, we find no discernible evidence of variability in either of the \textit{TESS} orbit bins, but future optical data on WASP-12b taken over longer timescales could reveal further evidence of this.

\section*{Acknowledgements}
 NO and CAW acknowledge support from Science and Technology Facilities Council (STFC) grants ST/T506369/1 and ST/P000312/1, respectively. M.J.H. acknowledges the support of the Swiss National Fund under grant 200020\_172746. This paper includes data collected by the TESS mission. This paper includes data collected with the TESS mission, obtained from the MAST data archive at the Space Telescope Science Institute (STScI). Funding for the TESS mission is provided by the NASA Explorer Program. STScI is operated by the Association of Universities for Research in Astronomy, Inc., under NASA contract NAS 5–26555. We also acknowledge use of the following Python packages: \texttt{SciPy} \citep{scipy}, \texttt{NumPy} \citep{numpy}, \texttt{emcee} \citep{emcee}, \texttt{Lightkurve} \citep{lightkurve}, \texttt{batman} \citep{batman}, \texttt{Matplotlib} \citep{matplotlib}, \texttt{AstroPy} \citep{astropy}.
 
 \section*{DATA AVAILABILITY}
The datasets used in this Letter are publicly available from the Mikulski Archive for Space Telescopes (MAST): https://archive.stsci.edu/.




\bibliographystyle{mnras}
\bibliography{main} 

\begin{thebibliography}{}
\makeatletter
\relax
\def\mn@urlcharsother{\let\do\@makeother \do\$\do\&\do\#\do\^\do\_\do\%\do\~}
\def\mn@doi{\begingroup\mn@urlcharsother \@ifnextchar [ {\mn@doi@}
  {\mn@doi@[]}}
\def\mn@doi@[#1]#2{\def\@tempa{#1}\ifx\@tempa\@empty \href
  {http://dx.doi.org/#2} {doi:#2}\else \href {http://dx.doi.org/#2} {#1}\fi
  \endgroup}
\def\mn@eprint#1#2{\mn@eprint@#1:#2::\@nil}
\def\mn@eprint@arXiv#1{\href {http://arxiv.org/abs/#1} {{\tt arXiv:#1}}}
\def\mn@eprint@dblp#1{\href {http://dblp.uni-trier.de/rec/bibtex/#1.xml}
  {dblp:#1}}
\def\mn@eprint@#1:#2:#3:#4\@nil{\def\@tempa {#1}\def\@tempb {#2}\def\@tempc
  {#3}\ifx \@tempc \@empty \let \@tempc \@tempb \let \@tempb \@tempa \fi \ifx
  \@tempb \@empty \def\@tempb {arXiv}\fi \@ifundefined
  {mn@eprint@\@tempb}{\@tempb:\@tempc}{\expandafter \expandafter \csname
  mn@eprint@\@tempb\endcsname \expandafter{\@tempc}}}

\bibitem[\protect\citeauthoryear{{Armstrong}, {de Mooij}, {Barstow}, {Osborn},
  {Blake}  \& {Saniee}}{{Armstrong} et~al.}{2016}]{armstrong}
{Armstrong} D.~J.,  {de Mooij} E.,  {Barstow} J.,  {Osborn} H.~P.,  {Blake} J.,
    {Saniee} N.~F.,  2016, \mn@doi [Nature Astronomy]
  {10.1038/s41550-016-0004}, \href
  {https://ui.adsabs.harvard.edu/abs/2016NatAs...1E...4A} {1, 0004}

\bibitem[\protect\citeauthoryear{{Astropy Collaboration} et~al.,}{{Astropy
  Collaboration} et~al.}{2018}]{astropy}
{Astropy Collaboration} et~al., 2018, \mn@doi [aj] {10.3847/1538-3881/aabc4f},
  \href {https://ui.adsabs.harvard.edu/abs/2018AJ....156..123A} {156, 123}

\bibitem[\protect\citeauthoryear{{Baraffe}, {Homeier}, {Allard}  \&
  {Chabrier}}{{Baraffe} et~al.}{2015}]{allard}
{Baraffe} I.,  {Homeier} D.,  {Allard} F.,   {Chabrier} G.,  2015, \mn@doi
  [\aap] {10.1051/0004-6361/201425481}, \href
  {https://ui.adsabs.harvard.edu/abs/2015A&A...577A..42B} {577, A42}

\bibitem[\protect\citeauthoryear{{Bell} et~al.,}{{Bell} et~al.}{2017}]{bell}
{Bell} T.~J.,  et~al., 2017, \mn@doi [\apjl] {10.3847/2041-8213/aa876c}, \href
  {https://ui.adsabs.harvard.edu/abs/2017ApJ...847L...2B} {847, L2}

\bibitem[\protect\citeauthoryear{{Bell} et~al.,}{{Bell} et~al.}{2019}]{bell2}
{Bell} T.~J.,  et~al., 2019, \mn@doi [\mnras] {10.1093/mnras/stz2018}, \href
  {https://ui.adsabs.harvard.edu/abs/2019MNRAS.489.1995B} {489, 1995}

\bibitem[\protect\citeauthoryear{{Bloemen} et~al.,}{{Bloemen}
  et~al.}{2011}]{doppler2}
{Bloemen} S.,  et~al., 2011, \mn@doi [\mnras]
  {10.1111/j.1365-2966.2010.17559.x}, \href
  {https://ui.adsabs.harvard.edu/abs/2011MNRAS.410.1787B} {410, 1787}

\bibitem[\protect\citeauthoryear{{Campo} et~al.,}{{Campo} et~al.}{2011}]{campo}
{Campo} C.~J.,  et~al., 2011, \mn@doi [\apj] {10.1088/0004-637X/727/2/125},
  \href {https://ui.adsabs.harvard.edu/abs/2011ApJ...727..125C} {727, 125}

\bibitem[\protect\citeauthoryear{{Claret}}{{Claret}}{2017}]{claret}
{Claret} A.,  2017, \mn@doi [\aap] {10.1051/0004-6361/201629705}, \href
  {https://ui.adsabs.harvard.edu/abs/2017A&A...600A..30C} {600, A30}

\bibitem[\protect\citeauthoryear{{Collins}, {Kielkopf}  \& {Stassun}}{{Collins}
  et~al.}{2017}]{collinsmass}
{Collins} K.~A.,  {Kielkopf} J.~F.,   {Stassun} K.~G.,  2017, \mn@doi [\aj]
  {10.3847/1538-3881/153/2/78}, \href
  {https://ui.adsabs.harvard.edu/abs/2017AJ....153...78C} {153, 78}

\bibitem[\protect\citeauthoryear{{Cowan}, {Machalek}, {Croll}, {Shekhtman},
  {Burrows}, {Deming}, {Greene}  \& {Hora}}{{Cowan} et~al.}{2012}]{cowan}
{Cowan} N.~B.,  {Machalek} P.,  {Croll} B.,  {Shekhtman} L.~M.,  {Burrows} A.,
  {Deming} D.,  {Greene} T.,   {Hora} J.~L.,  2012, \mn@doi [\apj]
  {10.1088/0004-637X/747/1/82}, \href
  {https://ui.adsabs.harvard.edu/abs/2012ApJ...747...82C} {747, 82}

\bibitem[\protect\citeauthoryear{{Croll}, {Lafreniere}, {Albert},
  {Jayawardhana}, {Fortney}  \& {Murray}}{{Croll} et~al.}{2011}]{croll}
{Croll} B.,  {Lafreniere} D.,  {Albert} L.,  {Jayawardhana} R.,  {Fortney}
  J.~J.,   {Murray} N.,  2011, \mn@doi [\aj] {10.1088/0004-6256/141/2/30},
  \href {https://ui.adsabs.harvard.edu/abs/2011AJ....141...30C} {141, 30}

\bibitem[\protect\citeauthoryear{{Crossfield}, {Barman}, {Hansen}, {Tanaka}  \&
  {Kodama}}{{Crossfield} et~al.}{2012}]{crossfield}
{Crossfield} I. J.~M.,  {Barman} T.,  {Hansen} B. M.~S.,  {Tanaka} I.,
  {Kodama} T.,  2012, \mn@doi [\apj] {10.1088/0004-637X/760/2/140}, \href
  {https://ui.adsabs.harvard.edu/abs/2012ApJ...760..140C} {760, 140}

\bibitem[\protect\citeauthoryear{{Delgado Mena} et~al.,}{{Delgado Mena}
  et~al.}{2015}]{stellarparams}
{Delgado Mena} E.,  et~al., 2015, \mn@doi [\aap] {10.1051/0004-6361/201425433},
  \href {https://ui.adsabs.harvard.edu/abs/2015A&A...576A..69D} {576, A69}

\bibitem[\protect\citeauthoryear{{Esteves}, {De Mooij}  \&
  {Jayawardhana}}{{Esteves} et~al.}{2013}]{esteves2013}
{Esteves} L.~J.,  {De Mooij} E. J.~W.,   {Jayawardhana} R.,  2013, \mn@doi
  [\apj] {10.1088/0004-637X/772/1/51}, \href
  {https://ui.adsabs.harvard.edu/abs/2013ApJ...772...51E} {772, 51}

\bibitem[\protect\citeauthoryear{{Esteves}, {De Mooij}  \&
  {Jayawardhana}}{{Esteves} et~al.}{2015}]{esteves2015}
{Esteves} L.~J.,  {De Mooij} E. J.~W.,   {Jayawardhana} R.,  2015, \mn@doi
  [\apj] {10.1088/0004-637X/804/2/150}, \href
  {https://ui.adsabs.harvard.edu/abs/2015ApJ...804..150E} {804, 150}

\bibitem[\protect\citeauthoryear{{F{\"o}hring}, {Dhillon}, {Madhusudhan},
  {Marsh}, {Copperwheat}, {Littlefair}  \& {Wilson}}{{F{\"o}hring}
  et~al.}{2013}]{foehring}
{F{\"o}hring} D.,  {Dhillon} V.~S.,  {Madhusudhan} N.,  {Marsh} T.~R.,
  {Copperwheat} C.~M.,  {Littlefair} S.~P.,   {Wilson} R.~W.,  2013, \mn@doi
  [\mnras] {10.1093/mnras/stt1443}, \href
  {https://ui.adsabs.harvard.edu/abs/2013MNRAS.435.2268F} {435, 2268}

\bibitem[\protect\citeauthoryear{{Foreman-Mackey}, {Hogg}, {Lang}  \&
  {Goodman}}{{Foreman-Mackey} et~al.}{2013}]{emcee}
{Foreman-Mackey} D.,  {Hogg} D.~W.,  {Lang} D.,   {Goodman} J.,  2013, \mn@doi
  [\pasp] {10.1086/670067}, \href
  {https://ui.adsabs.harvard.edu/abs/2013PASP..125..306F} {125, 306}

\bibitem[\protect\citeauthoryear{Harris et~al.,}{Harris et~al.}{2020}]{numpy}
Harris C.~R.,  et~al., 2020, \mn@doi [Nature] {10.1038/s41586-020-2649-2}, 585,
  357–362

\bibitem[\protect\citeauthoryear{{Hebb} et~al.,}{{Hebb} et~al.}{2009}]{wasp12b}
{Hebb} L.,  et~al., 2009, \mn@doi [\apj] {10.1088/0004-637X/693/2/1920}, \href
  {https://ui.adsabs.harvard.edu/abs/2009ApJ...693.1920H} {693, 1920}

\bibitem[\protect\citeauthoryear{{Hooton}, {de Mooij}, {Watson}, {Gibson},
  {Galindo-Guil}, {Clavero}  \& {Merritt}}{{Hooton} et~al.}{2019}]{hooton}
{Hooton} M.~J.,  {de Mooij} E. J.~W.,  {Watson} C.~A.,  {Gibson} N.~P.,
  {Galindo-Guil} F.~J.,  {Clavero} R.,   {Merritt} S.~R.,  2019, \mn@doi
  [\mnras] {10.1093/mnras/stz966}, \href
  {https://ui.adsabs.harvard.edu/abs/2019MNRAS.486.2397H} {486, 2397}

\bibitem[\protect\citeauthoryear{Hunter}{Hunter}{2007}]{matplotlib}
Hunter J.~D.,  2007, \mn@doi [Computing in Science \& Engineering]
  {10.1109/MCSE.2007.55}, 9, 90

\bibitem[\protect\citeauthoryear{{Jenkins} et~al.,}{{Jenkins}
  et~al.}{2016}]{spoc}
{Jenkins} J.~M.,  et~al., 2016, in Software and Cyberinfrastructure for
  Astronomy IV. p. 99133E, \mn@doi{10.1117/12.2233418}

\bibitem[\protect\citeauthoryear{{Kipping}}{{Kipping}}{2013}]{ldc}
{Kipping} D.~M.,  2013, \mn@doi [\mnras] {10.1093/mnras/stt1435}, \href
  {https://ui.adsabs.harvard.edu/abs/2013MNRAS.435.2152K} {435, 2152}

\bibitem[\protect\citeauthoryear{{Knutson} et~al.,}{{Knutson}
  et~al.}{2007}]{knutson2007}
{Knutson} H.~A.,  et~al., 2007, \mn@doi [\nat] {10.1038/nature05782}, \href
  {https://ui.adsabs.harvard.edu/abs/2007Natur.447..183K} {447, 183}

\bibitem[\protect\citeauthoryear{{Komacek}, {Showman}  \& {Tan}}{{Komacek}
  et~al.}{2017}]{contrast3}
{Komacek} T.~D.,  {Showman} A.~P.,   {Tan} X.,  2017, \mn@doi [\apj]
  {10.3847/1538-4357/835/2/198}, \href
  {https://ui.adsabs.harvard.edu/abs/2017ApJ...835..198K} {835, 198}

\bibitem[\protect\citeauthoryear{{Kreidberg}}{{Kreidberg}}{2015}]{batman}
{Kreidberg} L.,  2015, \mn@doi [\pasp] {10.1086/683602}, \href
  {https://ui.adsabs.harvard.edu/abs/2015PASP..127.1161K} {127, 1161}

\bibitem[\protect\citeauthoryear{{Lightkurve Collaboration}
  et~al.,}{{Lightkurve Collaboration} et~al.}{2018}]{lightkurve}
{Lightkurve Collaboration} et~al., 2018, {Lightkurve: Kepler and TESS time
  series analysis in Python}, Astrophysics Source Code Library (\mn@eprint
  {ascl} {1812.013})

\bibitem[\protect\citeauthoryear{{Loeb} \& {Gaudi}}{{Loeb} \&
  {Gaudi}}{2003}]{doppler1}
{Loeb} A.,  {Gaudi} B.~S.,  2003, \mn@doi [\apjl] {10.1086/375551}, \href
  {https://ui.adsabs.harvard.edu/abs/2003ApJ...588L.117L} {588, L117}

\bibitem[\protect\citeauthoryear{{L{\'o}pez-Morales} \&
  {Seager}}{{L{\'o}pez-Morales} \& {Seager}}{2007}]{seager}
{L{\'o}pez-Morales} M.,  {Seager} S.,  2007, \mn@doi [\apjl] {10.1086/522118},
  \href {https://ui.adsabs.harvard.edu/abs/2007ApJ...667L.191L} {667, L191}

\bibitem[\protect\citeauthoryear{{L{\'o}pez-Morales}, {Coughlin}, {Sing},
  {Burrows}, {Apai}, {Rogers}, {Spiegel}  \& {Adams}}{{L{\'o}pez-Morales}
  et~al.}{2010}]{lopes}
{L{\'o}pez-Morales} M.,  {Coughlin} J.~L.,  {Sing} D.~K.,  {Burrows} A.,
  {Apai} D.,  {Rogers} J.~C.,  {Spiegel} D.~S.,   {Adams} E.~R.,  2010, \mn@doi
  [\apjl] {10.1088/2041-8205/716/1/L36}, \href
  {https://ui.adsabs.harvard.edu/abs/2010ApJ...716L..36L} {716, L36}

\bibitem[\protect\citeauthoryear{{Mandel} \& {Agol}}{{Mandel} \&
  {Agol}}{2002}]{mandel}
{Mandel} K.,  {Agol} E.,  2002, \mn@doi [\apjl] {10.1086/345520}, \href
  {https://ui.adsabs.harvard.edu/abs/2002ApJ...580L.171M} {580, L171}

\bibitem[\protect\citeauthoryear{{Perna}, {Heng}  \& {Pont}}{{Perna}
  et~al.}{2012}]{contrast1}
{Perna} R.,  {Heng} K.,   {Pont} F.,  2012, \mn@doi [\apj]
  {10.1088/0004-637X/751/1/59}, \href
  {https://ui.adsabs.harvard.edu/abs/2012ApJ...751...59P} {751, 59}

\bibitem[\protect\citeauthoryear{{Showman} \& {Polvani}}{{Showman} \&
  {Polvani}}{2011}]{superrotation}
{Showman} A.~P.,  {Polvani} L.~M.,  2011, \mn@doi [\apj]
  {10.1088/0004-637X/738/1/71}, \href
  {https://ui.adsabs.harvard.edu/abs/2011ApJ...738...71S} {738, 71}

\bibitem[\protect\citeauthoryear{{Snellen}, {de Mooij}  \&
  {Albrecht}}{{Snellen} et~al.}{2009}]{snellencorot}
{Snellen} I. A.~G.,  {de Mooij} E. J.~W.,   {Albrecht} S.,  2009, \mn@doi
  [\nat] {10.1038/nature08045}, \href
  {https://ui.adsabs.harvard.edu/abs/2009Natur.459..543S} {459, 543}

\bibitem[\protect\citeauthoryear{{Stevenson}, {Bean}, {Madhusudhan}  \&
  {Harrington}}{{Stevenson} et~al.}{2014}]{stevenson}
{Stevenson} K.~B.,  {Bean} J.~L.,  {Madhusudhan} N.,   {Harrington} J.,  2014,
  \mn@doi [\apj] {10.1088/0004-637X/791/1/36}, \href
  {https://ui.adsabs.harvard.edu/abs/2014ApJ...791...36S} {791, 36}

\bibitem[\protect\citeauthoryear{{Turner}, {Ridden-Harper}  \&
  {Jayawardhana}}{{Turner} et~al.}{2020}]{turner2020}
{Turner} J.~D.,  {Ridden-Harper} A.,   {Jayawardhana} R.,  2020, arXiv
  e-prints, \href {https://ui.adsabs.harvard.edu/abs/2020arXiv201202211T} {p.
  arXiv:2012.02211}

\bibitem[\protect\citeauthoryear{Virtanen et~al.,}{Virtanen
  et~al.}{2020}]{scipy}
Virtanen P.,  et~al., 2020, \mn@doi [Nature Methods]
  {10.1038/s41592-019-0686-2}, \href {https://rdcu.be/b08Wh} {17, 261}

\bibitem[\protect\citeauthoryear{{Wong} et~al.,}{{Wong}
  et~al.}{2020}]{wongetal}
{Wong} I.,  et~al., 2020, \mn@doi [\aj] {10.3847/1538-3881/ababad}, \href
  {https://ui.adsabs.harvard.edu/abs/2020AJ....160..155W} {160, 155}

\bibitem[\protect\citeauthoryear{{Yee} et~al.,}{{Yee}
  et~al.}{2020}]{orbitdecay}
{Yee} S.~W.,  et~al., 2020, \mn@doi [\apjl] {10.3847/2041-8213/ab5c16}, \href
  {https://ui.adsabs.harvard.edu/abs/2020ApJ...888L...5Y} {888, L5}

\bibitem[\protect\citeauthoryear{{Zellem} et~al.,}{{Zellem}
  et~al.}{2014}]{infra1}
{Zellem} R.~T.,  et~al., 2014, \mn@doi [\apj] {10.1088/0004-637X/790/1/53},
  \href {https://ui.adsabs.harvard.edu/abs/2014ApJ...790...53Z} {790, 53}

\bibitem[\protect\citeauthoryear{{Zhang} et~al.,}{{Zhang}
  et~al.}{2018}]{zhang2018}
{Zhang} M.,  et~al., 2018, \mn@doi [\aj] {10.3847/1538-3881/aaa458}, \href
  {https://ui.adsabs.harvard.edu/abs/2018AJ....155...83Z} {155, 83}

\bibitem[\protect\citeauthoryear{{Zhao}, {Monnier}, {Swain}, {Barman}  \&
  {Hinkley}}{{Zhao} et~al.}{2012}]{zhao}
{Zhao} M.,  {Monnier} J.~D.,  {Swain} M.~R.,  {Barman} T.,   {Hinkley} S.,
  2012, \mn@doi [\apj] {10.1088/0004-637X/744/2/122}, \href
  {https://ui.adsabs.harvard.edu/abs/2012ApJ...744..122Z} {744, 122}

\bibitem[\protect\citeauthoryear{{de Mooij} \& {Snellen}}{{de Mooij} \&
  {Snellen}}{2009}]{demooij2009}
{de Mooij} E.~J.~W.,  {Snellen} I.~A.~G.,  2009, \mn@doi [\aap]
  {10.1051/0004-6361:200811239}, \href
  {https://ui.adsabs.harvard.edu/abs/2009A&A...493L..35D} {493, L35}

\bibitem[\protect\citeauthoryear{{von Essen} et~al.,}{{von Essen}
  et~al.}{2019}]{vonessen}
{von Essen} C.,  et~al., 2019, \mn@doi [\aap] {10.1051/0004-6361/201935312},
  \href {https://ui.adsabs.harvard.edu/abs/2019A&A...628A.115V} {628, A115}

\makeatother
\end{thebibliography}

\clearpage




\appendix

\section{Polynomial fitting and PDCSAP}\label{app:1}
\begin{table*}
	\centering
	\caption{Best-fit parameters for WASP-12b for PDCSAP flux, and SAP flux corrected via polynomial fitting. These parameters were found using an MCMC analysis similar to the approach used in Section \ref{analysis}}
	\label{tab:poly}
	\begin{tabular}{lccccr}
		\hline\hline
		Parameter & Prior & PDCSAP best-fit & Polynomial best-fit & TO1 (CBV) & TO2 (CBV)\\
		\hline
		\multicolumn{6}{c}{Orbital parameters} \\
		\hline
		$P$ (days) & fixed & \multicolumn{4}{c}{1.091420107}\\\
		\rule{0pt}{3.5ex}$t_{\mathrm{0}}$ (BJD) & fixed & \multicolumn{4}{c}{2456305.455809}\\
		\rule{0pt}{3.5ex}$dP/dN$ & fixed & \multicolumn{4}{c}{-10.04$\times$10$^{-10}$}\\
		\rule{0pt}{3.5ex}$a/R_{*}$ & $\mathcal{N}$(3.03, 0.034) & 3.038 $\pm$ 0.023 & 3.036 $\pm$ 0.024 & 3.047 $\pm$ 0.024 & 3.033$^{+0.023}_{-0.024}$\\
		\rule{0pt}{3.5ex}$i$ (degrees) & $\mathcal{N}$(83.37, 0.7) & 83.25$^{+0.50}_{-0.48}$ & 83.22$^{+0.55}_{-0.51}$ & 83.19$^{+0.53}_{-0.50}$ & 83.37$^{+0.53}_{-0.51}$\\
		\hline
		\multicolumn{6}{c}{Transit parameters} \\
		\hline
		$R_{\mathrm{p}}/R_{*}$ & - & 0.1173 $\pm$ 0.0007 & 0.1171 $\pm$ 0.0008 & 0.1181 $\pm$ 0.0009 & 0.1166 $\pm$ 0.0009\\
	    \rule{0pt}{3.5ex}$u_{1}$ & - & 0.212$^{+0.159}_{-0.114}$ & 0.162$^{+0.165}_{-0.106}$ & 0.314$^{+0.259}_{-0.176}$ & 0.146$^{+0.173}_{-0.102}$\\
	    \rule{0pt}{3.5ex}$u_{2}$ & - & 0.301$^{+0.179}_{-0.132}$ & 0.374$^{+0.190}_{-0.135}$ & 0.146$^{+0.178}_{-0.261}$ & 0.405$^{+0.203}_{-0.137}$\\
	    \hline
	    \multicolumn{6}{c}{Phase curve} \\
	    \hline
	    $F_{\mathrm{ecl}}$ (ppm) & - & 577$^{+71}_{-72}$ & 494$^{+81}_{-83}$ & 655$^{+96}_{-98}$ & 553 $\pm$ 110\\
	    \rule{0pt}{3.5ex}$A_{\mathrm{p}}$ (ppm) & - & 530 $\pm$ 60 & 537 $\pm$ 73 & 530$^{+80}_{-82}$ & 563$^{+92}_{-94}$\\
	    \rule{0pt}{3.5ex}$\theta$ (phase) & - & 0.061 $\pm$ 0.015 & -0.002 $\pm$ 0.017 & 0.063 $\pm$ 0.022 & 0.034 $\pm$ 0.021\\
	    \rule{0pt}{3.5ex}$M_{\mathrm{p}}$ (M$_{\mathrm{J}}$) & $\mathcal{N}$(1.47, 0.076) & 1.48 $\pm$ 0.08 & 1.48 $\pm$ 0.08 & 1.48 $\pm$ 0.08 & 1.47 $\pm$ 0.08\\
	    \rule{0pt}{3.5ex}$F_{\mathrm{n}}$ (ppm) & - & 47 $\pm$ 94 & -43 $\pm$ 111 & 125$^{+125}_{-128}$ & -10$^{+144}_{-145}$\\
		\hline
	\end{tabular}
\end{table*}

For completeness, we jointly fit the phase curve model with an 11th order polynomial for \textit{TESS} Orbit 1 (TO1) and a 5th order polynomial for \textit{TESS} Orbit 2 (TO2), using a similar strategy as outlined in in Sect.~\ref{sec:phasecurve} of the form:
\begin{equation}
    P_{n}(x) = \sum_{k=0}^{n} a_{k} x^{k}
	\label{eq:polynomial}
\end{equation}
while sharing parameters specific to the stellar-planetary system. The polynomial order was found by calculating and minimising the Bayesian Information Criterion (BIC) for TO1 and TO2 independently of one another. However, the system parameters derived from each fit differed significantly between both orbits, suggesting that the systematics present are on shorter time-scales than initially thought. Higher order fits prove fruitless as any further polynomial terms could risk fitting and jeopardising the derived phase curve parameters. As a result of these discrepancies, we opt not to use these polynomial fits in deriving our final planetary parameters.

We also include here the individual fits for each \textit{TESS} orbit using CBVs to detrend the series. We find no significant variability among the binned orbits; however, it must be noted that by binning over each orbit, any variations on short time-scales will be averaged out. A search for variability would require multiple follow-ups of the same system in the same bandpass at significantly different times. Simultaneous observations of WASP-12b could also help rule out the effect of systematics on the observed eclipse depth. We also present here fits for the PDCSAP flux which has been thoroughly detrended using SPOC algorithms. We find the results of these fits to be compatible with the main fit performed in Section \ref{analysis}. Table \ref{tab:poly} show our best fit parameters for each dataset and fit.

\section{Additional plots}\label{app:2}

We also present here the corner plot for our fit in Section \ref{analysis} as well as Figure \ref{fig:figure2} fixated on the transit to better identify the quality of the different ephemeris fits.

\begin{figure*}
	\includegraphics[width=\textwidth]{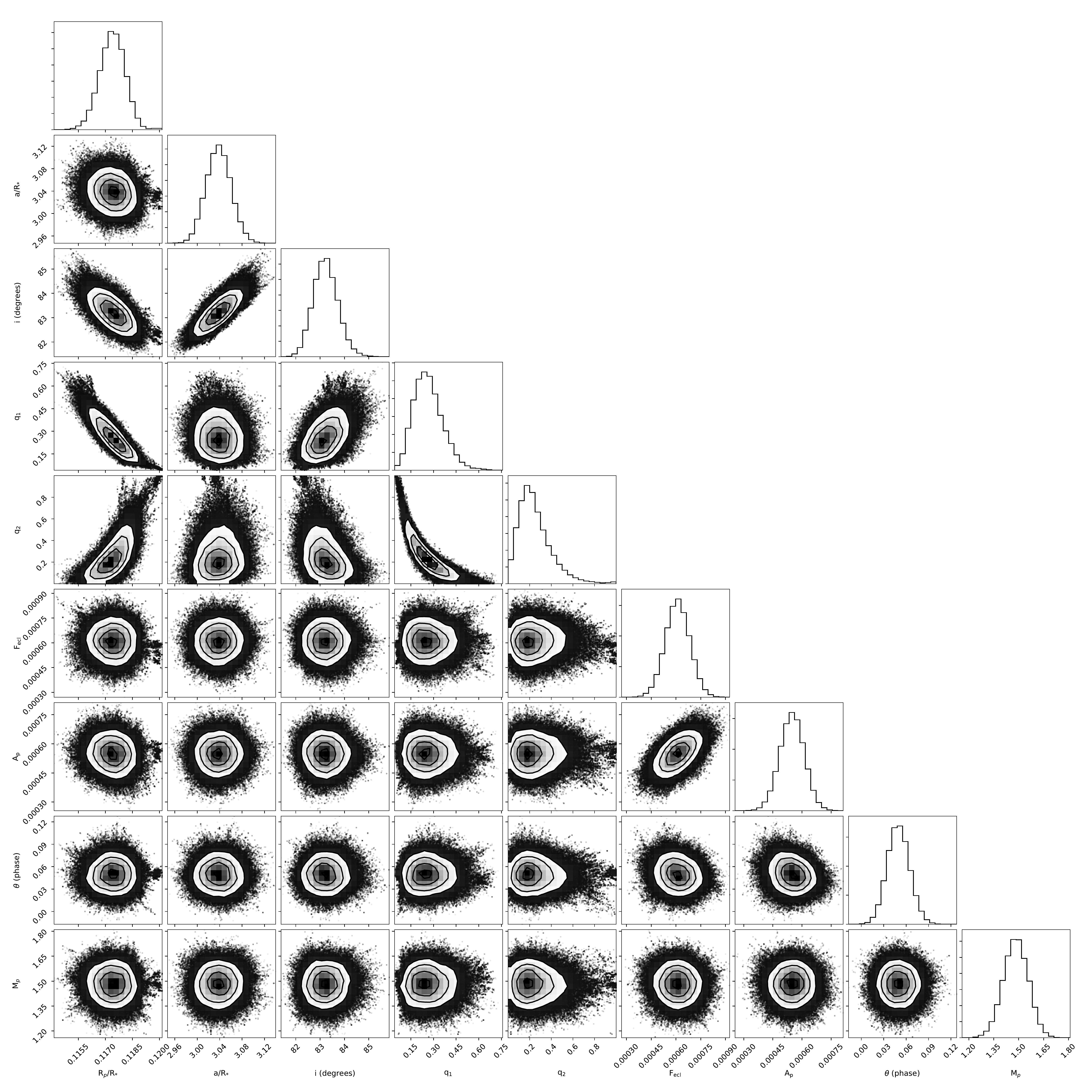}
    \caption{Reduced size corner plot from our analysis in Section \ref{analysis}.}
\end{figure*}

\begin{figure*}
	\includegraphics[width=\textwidth]{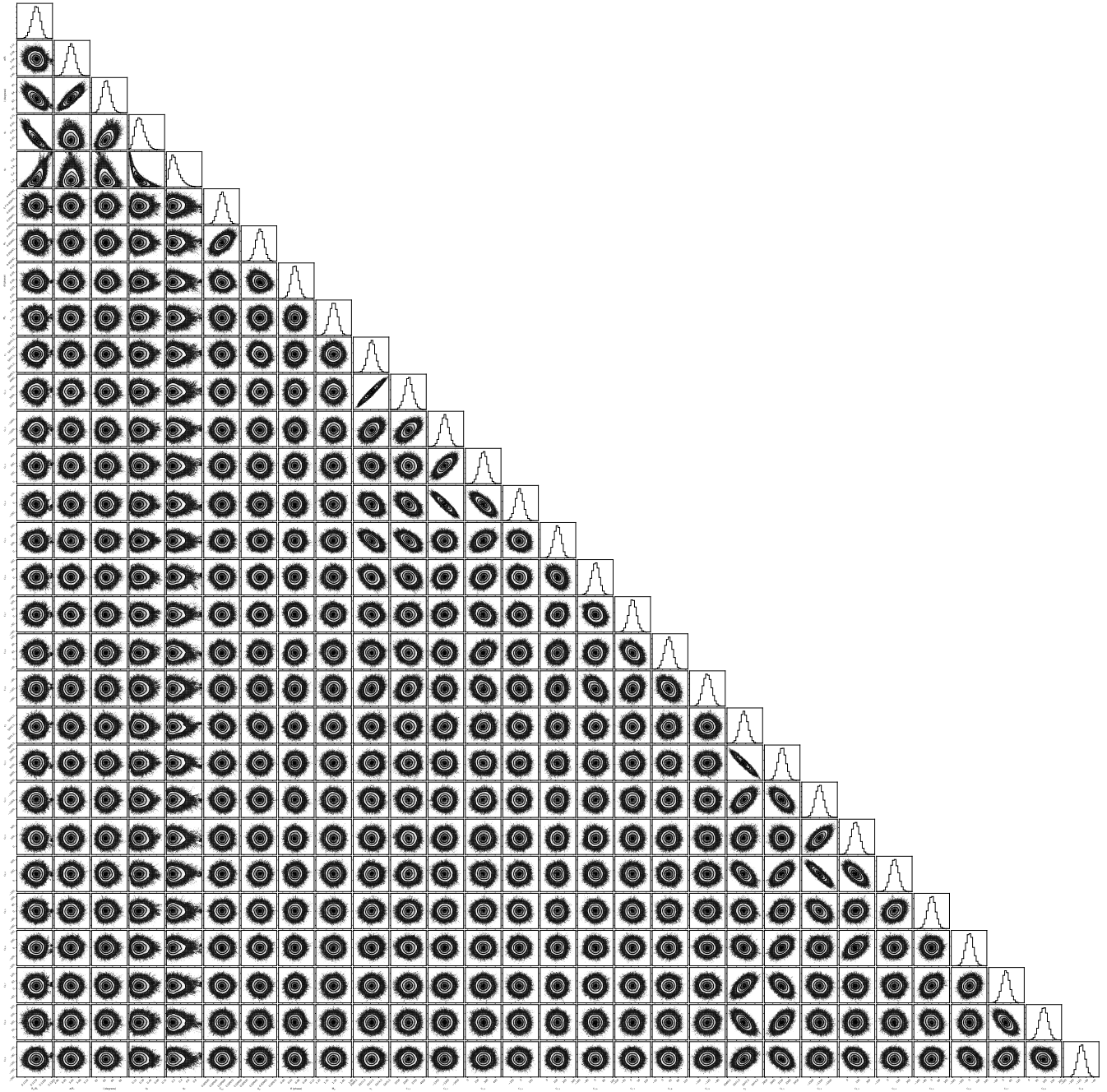}
    \caption{Full corner plot from our analysis in Section \ref{analysis}, including coefficients for CBVs (included to show that CBV coefficients are not degenerate with planetary/transit parameters).}
\end{figure*}

\begin{figure*}
	\includegraphics[width=\textwidth]{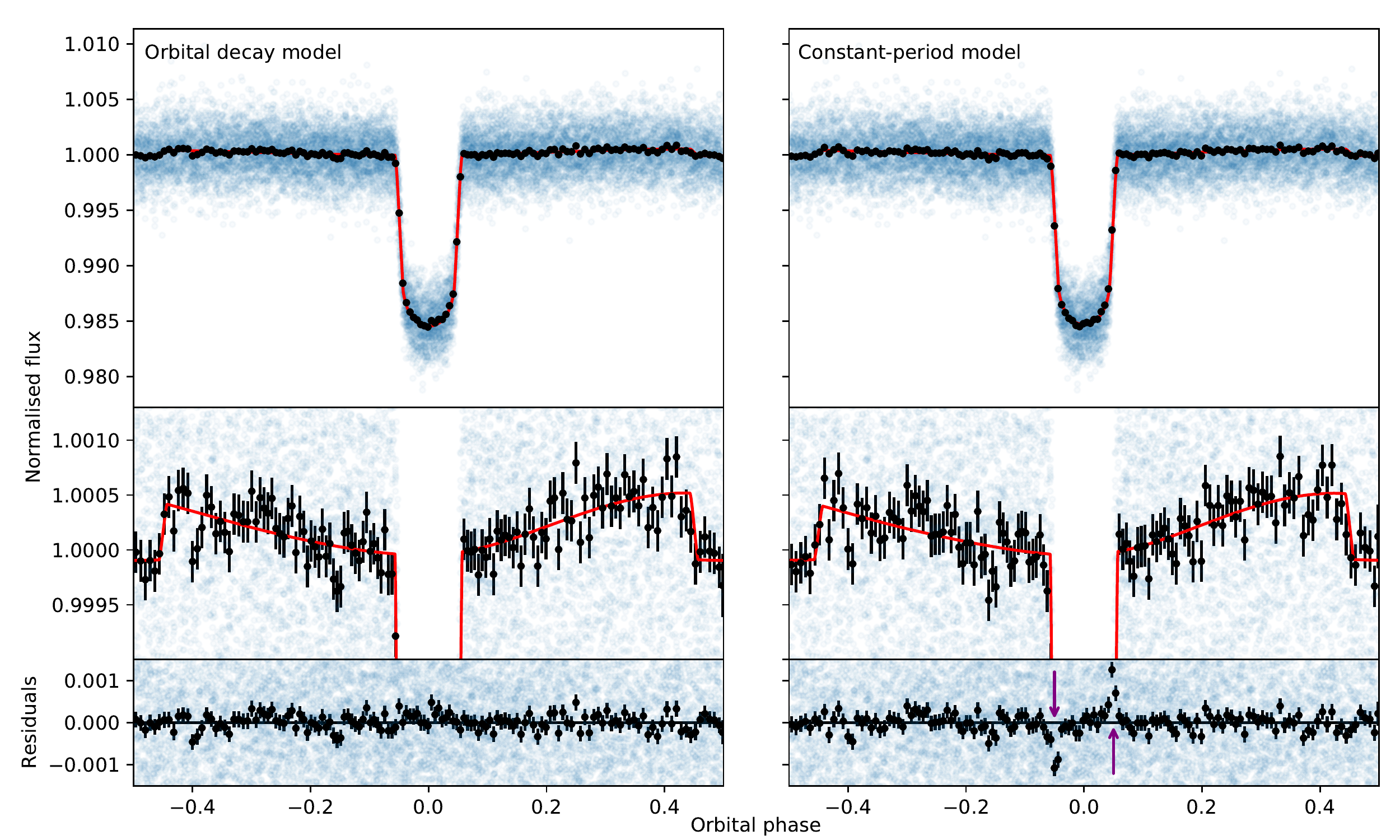}
    \caption{Same as Figure \ref{fig:figure2}, but centred on the transit at $\phi$ = 0. Residual artefacts are clearly present in the constant-period model plot.}
\end{figure*}

\bsp	
\label{lastpage}
\end{document}